\documentclass[onecolumn, 12pt]{ieeetran}

\newcounter{chapter}
\usepackage{epsfig,amssymb}
\usepackage{graphics,color}
\usepackage{amsmath,amssymb,amsfonts,amscd}
\usepackage[boxed,algochapter,noline,czech]{algorithm2e}

\newcommand{\be}{\begin{equation}}
\newcommand{\ee}{\end{equation}}
\newcommand{\ba}{\begin{array}}
\newcommand{\ea}{\end{array}}
\newcommand{\bea}{\begin{eqnarray}}
\newcommand{\eea}{\end{eqnarray}}

\newcommand{\VVor}{\mathcal{V}(\mathcal{R})}

\newcommand{\mR}{\mathcal{R}}
\newcommand{\NSM}{G_n(\mR)}
\newcommand{\bfemf}[1]{\ensuremath{\boldsymbol{#1}}}
\renewcommand{\vec}[1]{\ensuremath{\boldsymbol{#1}}}
\newcommand{\lmb}{\vec{\lambda}}

\newtheorem{theorem}{Theorem}
\newtheorem{corollary}{Corollary}

\begin{document}

\title{ Linear code-based vector quantization for independent
random variables}

\author{\authorblockN{\hspace*{-2mm}Boris
D. Kudryashov}\\
 \authorblockA{\hspace*{-2mm} Department of Information Systems,\\
\hspace*{-2mm} St. Petersburg Univ. of Information Technologies, Mechanics and Optics\\
\hspace*{-2mm}St. Petersburg 197101, Russia  \\
\hspace*{-2mm}Email: boris@eit.lth.se} \and\\
\authorblockN{\hspace*{-2mm} Kirill V. Yurkov}\\
\authorblockA{Department of Information Systems,\\
 St. Petersburg University on Aerospace Instrumentation, \\
 St. Petersburg, 190000, Russia\\
Email: yourkovkirill@mail.ru} }

\maketitle

\begin{abstract}
 The computationally efficient vector quantization for a
discrete time source can be performed using lattices over block
linear codes or convolutional codes. For high rates (low
distortion) and uniform distribution
the performance of a multidimensional lattice depends
mainly on the normalized second moment (NSM) of the lattice.
For relatively low rates (high distortions) and non-uniform distributions
the lattice-based quantization can be
non-optimal in terms of achieving the Shannon rate-distortion function
$H(D)$.

In this paper we analyze the rate-distortion function $R(D)$
achievable using linear codes over $GF(q)$, where $q$ is a prime
number. We show that even for $q=2$ the NSM of code-based quantizers
is close to the minimum achievable limit.
If $q \to \infty$, then NSM $\to 1/(2\pi e)$ which is the NSM of the
infinite dimension sphere. By
exhaustive search over $q$-ary time-invariant convolutional codes
with memory $\nu\le 8$, the NSM-optimum codes for $q=2,3,5$ were found.

For low rates (high distortions) we use a non-lattice quantizer
obtained from a lattice quantizer  by extending the ``zero zone''.
Furthermore, we modify the encoding by a properly assigned metric to
approximation values. Due to such modifications
for a wide class of probability distributions and
a wide range of bit rates we obtained up to
1 dB signal-to-noise ratio improvement compared to currently known
results.

\end{abstract}
{\bf Index terms}---Arithmetic coding, convolutional codes, lattice,
lattice quantization, normalized second moment, random coding, rate-distortion function, scalar quantization, vector quantization
\section{Introduction}

For an arbitrary discrete time
memoryless continuous stationary source $X=\{x\}$, approximation alphabet
$Y=\{y\}$, and a given distortion measure $d(x,y)$ the theoretical limit
of the code rate $R$ under the condition that the average distortion
does not exceed $D$
is determined by the Shannon rate-distortion
function $H(D)$ \cite{Cover&Thomas_1991},
\[
H(D)=\min_{\{f(y|x)\}: {\rm E} [d(x,y)]\le D} \{I(X;Y)\}
\]
where $I(X;Y)$ denotes the mutual information between $X$ and $Y$.

An encoder of a discrete time continuous source is often referred
to as a vector quantizer, and the code as a codebook \cite{Gray_1998}. Coding
theorems about achievability of $H(D)$ are proved by random coding
techniques without imposing any restrictions on the codebook structure.
The encoding complexity for such a codebook in general is
proportional to the codebook size; i.e., the encoding
complexity for sequences of length $n$ with rate $R$ bits per
sample is proportional to $2^{Rn}$.

A large step towards reducing the quantization complexity is due
to using quantizers based on multidimensional lattices
\cite{Conway&Sloane_1993}. In particular, Conway and Sloane in
\cite{Conway&Sloane_1982_1} and \cite{Conway&Sloane_1982_2}
proposed a simple encoding algorithm,
which, in essence, requires a proportional to dimension $n$ of
a codebook number of scalar quantization operations and subsequent
exhaustive search over the lattice points in vicinity of the source
sequence. The complexity of this quantization still grows
exponentially in the dimension $n$ of the vectors, but much slower
than $2^{Rn}$.

Important results related to the vector quantization
efficiency are obtained by Zador \cite{Zador_1963}, who has shown that,
in the case of small quantization errors, the characteristics of the
quantizer depend on a parameter $G_n$ which is determined
by the shape of the Voronoi polyhedrons of the codebook
entries. For lattice quantizers, i.e., when
all Voronoi regions are congruent to each other, $G_n$ is equal to the
so-called \emph{normalized second moment} (NSM) of the lattice (see
(\ref{NSMdef}) in Section 2 below).

Constructions of lattices with good values of NSMs are presented
in \cite{Conway&Sloane_1993}. For example, for the Leech lattice
$\Lambda_{24}$ NSM = 0.0658. Since for lattice $\mathbb{Z}$ (the
set of integers) NSM = 1/12, the so called ``\emph{granular gain}''
of a lattice quantizer in dB is computed as $10\log_{10} (12 \rm{NSM})$.
Therefore the granular gain of $\Lambda_{24}$ with respect to $\mathbb{Z}$
is equal to 1.029 dB which is 0.513 dB away from the theoretical
limit $10 \log_{10} (12/2\pi e)=1.54$ dB corresponding to covering the Euclidean space by $n$-dimensional spheres when $n\to \infty$.

Granular gains achieved by different code constructions were
reported in \cite{Forney&Eyuboglu_1993}. In particular, the Marcellin-Fischer
\cite{Marcellin&Fischer_1990} trellis-based quantizers are
presented in \cite{Forney&Eyuboglu_1993} as an NSM record-holder. For example, by a trellis
with 256 states at each level MSN=1.36 dB is achieved (only
$\approx$0.17 dB from the sphere-covering bound).

The Marcellin-Fisher \cite{Marcellin&Fischer_1990} quantizer nowadays
is still the best-known in the sense of granular gain. It means
that these quantizers are good for uniformly distributed random
variables. Meanwhile results reported in
\cite{Marcellin&Fischer_1990} for the Laplacian and Gaussian
distributions were later improved in the papers \cite{FisherWang_92},
\cite{LarFarvard94}, \cite{Marcel94}, and
\cite{VleuWeb95}. Furthermore, the best currently known results for
the generalized Gaussian distribution with parameter $\alpha =0.5$
is presented in  \cite{Fischer&Yang_1998}.


%

Despite numerous attempts to construct good quantization schemes,
the gap between asymptotically achievable
results and the best existing quantizers is still large.
For example, for 32-state trellises and bit rate 3 bits/sample,
this gap is equal to 1.10, 1.53, and 1.78 dB for the Gaussian distribution,
Laplacian distribution, and
generalized Gaussian distribution with $\alpha =0.5$, respectively
\cite{Fischer&Yang_1998}.

Since the best codes and coding algorithms for non-uniform
distributions are not the same as for the uniform distribution, the following
question arises: could it be that codes which deliberately are
not good in sense of granular gain be very good for a non-uniform
distribution?

The positive answer to this question is one of the results of this paper.
Below we present the coding scheme which with a 32-state trellis at 3 bits/sample
is not more than 0.32, 0.35, and 0.44 dB away from the Shannon limit for the three
distributions mentioned above. Therefore, the achieved gain over existing
trellis quantizers  exceeds 1 dB
at high rates for a wide class of probability distributions.

We have searched good quantizers among lattice quantizers
over linear codes. The motivation is
that the constructions of some good lattices are closely
related to the constructions of good block error-correcting codes
(e.g., the Hamming code and Golay codes).
One more argument in favor of such trellises appears in
\cite{Kudryashov&Yurkov_PIT_2007}, \cite{Kudryashov&Yurkov_ISIT_2007} and
\cite{Litsyn_2005},
where it was shown that there exist asymptotically optimal lattices
over $q$-ary linear codes as $q$ tends to infinity.

Although the NSM decreases (the granular gain increases) with alphabet size $q$,
it was found by simulations that the quantization performance behavior
for non-uniform quantization is different: binary codes appear to be
better than codes over larger fields.
Replacing the quaternary codes, typically considered the
in trellis-based quantization literature,
by binary codes is the first step towards
improving the quantization performance.

Another step is based on the important observation
(see, e.g., \cite{Farvardin&Modestino_1984}, \cite{KudrPorovOh})
that the entropy-coded uniform
scalar quantization with properly chosen reconstruction values
provides near the same coding efficiency as the optimum entropy-constrained
scalar quantization (ECSQ) does.
Moreover, choosing zero quant larger than others, we obtain a simple scalar
quantizer with performance extremely close to the ECSQ for all rates.
In \cite{KudrPorovOh} this quantizer is called Extending Zero Zone (EZZ)
quantizer. Notice that EZZ quantization does not follow the minimum Euclidean
distance criterion. Therefore, it would be reasonable not to use the
Euclidean distance for vector
quantization (or trellis quantization as a special case) as well.

We begin in Section 2 with reviewing the random coding bounds
\cite{Kudryashov&Yurkov_PIT_2007}, \cite{Kudryashov&Yurkov_ISIT_2007}
on the NSM value for linear code based lattices.
The asymptotic behavior of the NSM as a  function of the code
alphabet size is studied and we show that code-based lattices are
asymptotically optimal when the alphabet size tends to infinity.
In Section 3 the code search results are presented.
In Section 4 we present entropy-constrained code-based
quantization. By generalizing the EZZ approach to lattice quantization, we
obtained near optimum SNR values for all code rates for the classes
of non-uniform probability distributions mentioned above. Concluding remarks are given in Section 5.

\section{Code-based quantization}

An $n$-dimensional lattice can be defined by its generator matrix
$G\in \mathbb{R}^{n\times n}$ as \cite{Conway&Sloane_1993}
\begin{equation}
\label{Lattice def1}
\Lambda(G)=\{\lmb \in \mathbb{R}^n|\exists \bfemf{z}
\in \mathbb{Z}^n:\lmb = G\bfemf{z}\}.
\end{equation}

Let $q$ be a prime number and $C$ denote $q$-ary linear $(n,k)$ code.
Then a lattice over the code $C$ is defined as
\begin{equation}
\label{Lattice def2}
\Lambda(C)=\{\lmb \in \mathbb{Z}^n|\exists \bfemf{c}\in C:
\lmb \equiv \bfemf{c}(\mbox{mod} \quad q)\}.
\end{equation}
In other words, the lattice $\Lambda(C)$ consists of integer-valued
sequences which are equal to codewords from $C$ by modulo $q$.
In the case of binary codes the least significant bits of the sequences
$\lmb \in \Lambda(C)$ are codewords of $C$.

The lattice defined by (\ref{Lattice def2}) is known as Construction A
\cite{Conway&Sloane_1993}.

When a $n$-dimensional lattice is used as a codebook, the source sequence
$\bfemf{x}\in \mathbb{R}^n$ is approximated by some lattice
point $\lmb\in \Lambda(C)$. The whole set $\Lambda(C)$ is assumed to be
ordered
and a number representing the optimal approximation point $\lmb$ is transmitted
instead of $\bfemf{x}$. Therefore, we need some rule which assigns
a number to each point of $\Lambda(C)$. To suggest such a rule we
consider the following representation of $\Lambda(C)$:

\begin{equation}
\label{lattice_decomposition_cosets}
\Lambda(C)=\bigcup_{m=1}^{q^k} \{q\mathbb{Z}^n+\vec{c_m}\}.
\end{equation}
In other words, $\Lambda(C)$ is a union of $q^k$ cosets which
leaders are coderwords of $C$. This representation allows a one-to-one
mapping of all $\lmb\in \Lambda(C)$ onto the set of integer vectors
$(m,\bfemf{b})$, where $m$ is a codeword number and
$\bfemf{b}=(b_1,...,b_n)\in \mathbb{Z}_n$ defines the
point number in $q\mathbb{Z}^n$. The lattice point
$\lmb$ can be reconstructed from $(m,\bfemf{b})$ by the
formula
\begin{equation}
\lmb=\vec{c_m}+q\vec b.
\label{lattice point}
\end{equation}

The lattice $\Lambda(C)$ or its scaled version $a\Lambda(C), a>0$
can be used as a codebook for vector quantization.
Below we always assume $a=1$ since instead of quantizing the random
vector $\vec x$ with $a\Lambda(C)$ we can quantize $\vec x/a$ with
$\Lambda(C)$. We assume also that
\begin{itemize}
    \item
     The stationary memoryless source sequences are drawn from
$X^n=\{\bfemf{x}|\bfemf{x}=(x_1,\ldots ,x_n)\}\subseteq
\mathbb{R}^n$ with a known probability density function (pdf)
$f(\bfemf{x})=\prod_{i=1}^{n} f(x_i)$.
   \item
     The squared Euclidean distance
     \[
     d(\bfemf{x},\bfemf{y})=\frac{1}{n}\| \bfemf{x}-\bfemf{y}\|^2
     =\frac{1}{n}\sum_{i=1}^{n} (x_i-y_i)^2
     \]
     is used as a distortion measure.
\end{itemize}

The quantization procedure can be described as a mapping $\bfemf{y}=Q(\bfemf{x})$,
where $\bfemf{x}\in X^n$, $\bfemf{y}\in \Lambda(C)$.
The mean square error (MSE) of the quantizer
\begin{equation} \label{gDistortion}
 D=\frac{1}{n}M\left[|| \bfemf{x} - Q(\bfemf{x}) || ^2\right]=
 \frac{1}{n}\sum_{i}\int_{\mathcal{R}_i}|| \bfemf{x} -
\lmb_i || ^2 f(\bfemf{x})d\bfemf{x}
\end{equation}
is minimized by the mapping
\[
Q(\bfemf{x})=\lmb_i \mbox{  if  } \vec{x}\in \mathcal{R}_i
\]
where $\mathcal{R}_i$ are Voronoi regions \cite{Conway&Sloane_1993}  of lattice points, i.e.,
$$
\mathcal{R}_i
\triangleq \mathcal{R}(\lmb_i) = \{\bfemf{x}\in
\mathbb{R}^n:||\bfemf{x}-\lmb_i||^2 \leq ||\bfemf{x}-\lmb_j||^2
\quad \forall j\neq i\}.
$$
All Voronoi regions of lattice points are congruent to each other. We denote the Voronoi region which contains the origin as $\mathcal{R}$.
We use the conventional notations
\begin{eqnarray}
\VVor&=& \int_{\mathcal{R}}d\bfemf{x} \nonumber \\
\NSM &=& \frac{1}{n} \frac{\int_{\mathcal{R}} \| \bfemf{x}\|^2 d\bfemf{x} }{\VVor^{1+2/n}} \label{NSMdef}
\end{eqnarray}
for the volume and the normalized second moment (NSM) of $\mathcal{R}$, respectively.

By entropy coding the rate $R$ of the lattice quantizer can be arbitrary close to the entropy of the lattice points
$$
R=-\frac{1}{n}\sum_{i}p_i\log{p_i},
$$
where the probabilities $p_i$ of the lattice points are equal to
$$
 p_i \triangleq p(\lmb_i)=\int_{\mathcal{R}_i}f(\bfemf{x})d\bfemf{x},
 \quad i=1,2,\ldots.
$$

The rate-distortion function $R(D)$ of the lattice quantizer can be easily
estimated under the assumption that the quantization errors are so small
that the pdf  $f(\bfemf{x})$ can be approximated by a constant value
inside each region $\mathcal{R}(\lmb_i)$. In this case, the rate
can be expressed via the differential entropy of the source
\[
h(X)=- \int  f(x) \log f(x) dx
\]
and the volume $\VVor$ of the Voronoi region $\mathcal{R}$,
\begin{equation}\label{RviaV}
R(D) \leq h(X)-\frac{1}{n}\log\VVor.
\end{equation}
For the average distortion value $D$ from (\ref{gDistortion}) and (\ref{NSMdef}) we have
\begin{equation}
\label{Distortion-SecMoment}
D = G_n(\mR)\VVor^{2/n}.
\end{equation}
From the last two equations and the Shannon lower bound,
\[
H(D)\ge h(X)-\frac{1}{2}\log(2\pi e D)
\]
we obtain the inequalities
\begin{equation}
\label{ZadorRD}
H(D) \leq R(D) \leq H(D)+\frac{1}{2}\log{(2\pi e \NSM)}.
\end{equation}

Therefore to prove that the optimal quantization
would be asymptotically achievable by lattice quantization
(i.e., $R(D) \to H(D) $, if $n\to \infty)$ we need to prove that
$ \NSM \to 1/(2\pi e) $ when $n\to \infty$.

Consider a lattice  $\Lambda(C) $ over a $q$-ary  linear
($n,k$) code, and denote the code rate $R_{\rm C} =k/n$. It can
be shown (see \cite{Kudryashov&Yurkov_PIT_2007}) that the overall volume of the $q^k$ Voronoi regions
of the codewords of $C$ coincides with the volume of the $n$-dimensional cube
with edge length equal to $q$. It means that $\VVor=q^{n}/q^{k}=q^{n-k}$ and from
(\ref{Distortion-SecMoment}) follows
\begin{equation}
\label{DistortionInHyperCube} D=\NSM q^{2(1-R_{\rm C})}.
\end{equation}

Formally we cannot use equation (\ref{DistortionInHyperCube}) for finding
NSM for a lattice over a code by estimating the quantization error for a random variable uniformly
distributed over the cube. The reason is that the $n$-dimensional cube
with length of the edge $q$ does not contain precisely $q^k$ complete Voronoi regions.
Some regions near the boundaries of the cube protrude from the cube.
To solve this problem the ``cyclic'' metric
\begin{equation} \label{rho}
\rho(x,y) = \min\{(x-y)^2,(|x-y|-q)^2\}
\end{equation}
was introduced in \cite{Kudryashov&Yurkov_PIT_2007}. It was shown that
the NSM of the lattice satisfies (\ref{DistortionInHyperCube}) if
the lattice points of the cube are used as codebook entries and the distortion $D$ is computed
with respect to the metric $\rho(\cdot,\cdot)$.

Below we will use (\ref{DistortionInHyperCube}) to
estimate the NSM of convolutional codes and in \cite{Kudryashov&Yurkov_PIT_2007}
this equation was used to obtain the random coding
bound on the achievable NSM value. The reformulated result of
\cite{Kudryashov&Yurkov_PIT_2007} (see comments in Appendix) is presented here
without a proof.

\begin{theorem}
\label{Main Teorem}
For any prime number $q$ and any $\varepsilon > 0$, there exists a
sequence $C_1,C_2,C_3\ldots$, of $q\,$-ary linear codes, where $C_n$
is a code of length $n$, and such a number $N$, that for all $n\geq N$
\begin{equation}
\label{Final inequality on second moment}
|G_n(q)-d_0q^{2(R_0(q)-1)}|<\varepsilon
\end{equation}
where $G_n(q)$ is the second normalized moment of the lattice based
on the code $C_n$, $d_0$ is a solution of the equation
\begin{equation}
\label{d_0 calculation}
d=2\left.\int_0^{1/2} \frac{\displaystyle \frac{\partial}
{\partial s}g(s,x)}{g(s,x)}dx\right|_{s=-\frac{1}{2d}}
\end{equation}
and $R_0(q)$ is defined as
\begin{equation}
\label{R_0 calculation}
R_0(q)=\frac{1}{\ln q}
\Bigl(
-\frac{1}{2}-2\int_0^{1/2}
\ln g\left( 1/(2d_0),x \right) dx
\Bigr)
\end{equation}
where
\begin{equation}
\label{gsx}
g(s,x)=\frac{1}{q}\sum_{k=0}^{q-1}e^{s\rho(x,k)}
\end{equation}
and $\rho(\cdot,\cdot)$ is defined by (\ref{rho}).
\end{theorem}

\begin{table}
\label{opt_nsm_table}
\caption{Random coding upper bounds on NSM for optimum code rate and for
code rate $R_{\rm C}=1/2$}

\begin{center}
\begin{tabular}{|c|c|c|c|}
  \hline
 $q$     & $R_0$  & \multicolumn{2}{c|}{$G_\infty(q)$} \\  \cline{3-4}
          &        & $R_C=R_0$&$R_{\rm C}=1/2$\\  \hline
  2       & 0.4144 & 0.0598 & 0.0631 \\  \hline
  3       & 0.4633 & 0.0587 & 0.0592 \\  \hline
  5       & 0.5000 & 0.0586 & 0.0586 \\  \hline
  7       & 0.5000 & 0.0585 & 0.0585 \\  \hline
  $\infty$& 0.5000 & 0.0585 & 0.0585 \\  \hline
\end{tabular}

\end{center}
\end{table}

The values $G_\infty(q)=\lim_{n\to \infty} G_n(q)$ for $q=2,3,5,7$ are given in Table \ref{opt_nsm_table}.
Also we show in the table the code rate value $R_{\rm C}=R_0(q)$ which minimizes the
random coding estimate on the NSM value $G_\infty(q)$ and the NSM value achievable when
we consider codes with nonoptimal rate  $R_{\rm C}=1/2$.
Presented estimates give the achievable NSM values if long enough codes
are used for quantization. In particular, the search results presented in Section
\ref{NSMsection} show that the lattices obtained from convolutional codes with a
number of encoder states above 512 are roughly 0.2 dB away from the asymptotically achievable
NSM value for a given alphabet size $q$.
On the other hand, rate $R_{\rm C}=1/2$ binary codes with 64 states have even better NSM
than the NSM computed
by averaging over randomly chosen infinitely long rate $R_{\rm C}=1/2$ codes.

An analysis of the asymptotic behavior of $G_\infty(q)$ and $R_0(q)$ with $q\to \infty$ leads to the
following result:

\begin{corollary}
\label{Cor_1}
\begin{eqnarray}
\lim_{q \to \infty } G_\infty(q) &=& \frac{1}{2\pi e} \label{Ginf}\\
\lim_{q \to \infty } R_0(q) &=& \frac{1}{2}. \label{Rinf}
\end{eqnarray}

\end{corollary}
The proof is given in Appendix.

\section{NSM-optimum convolutional codes}
\label{NSMsection}

It follows from Corollary 1 that codes with rate $R_{\rm C}=1/2$ provide
asymptotically optimum NSM value for large code alphabet size $q$.
It follows from the data in Table \ref{opt_nsm_table} that codes with rate $R_C=1/2$
give near-optimum NSM except for $q=2$.
Below we consider only rate $R_C=1/2$ codes since they are
most convenient for practical applications.

Moreover, we will search for good linear codes for quantization among truncated
convolutional codes, since for such codes the quantization (search for a codeword closest
to the source sequence) can be easily performed
using the Viterbi algorithm. 

A $q$-ary, rate $R_C=1/2$ convolutional encoder
over $GF(q)$ with memory $m$ can be described by the polynomial encoding
matrix
$$
G(D)=\left ( g_1(D) \quad g_2(D) \right)
$$
where
\begin{eqnarray*}
g_{i}(D)=g_{i0}+g_{i1} D+ \cdots+ g_{im}D^m,\quad
i=1,2
\end{eqnarray*}
are $q$-ary polynomials such that $\max_{i}\left\{\deg {g}_{i}(D)\right\}=m$.

In polynomial notations the infinite length input information sequence
$$
u(D) = u_0 + u_1 D + \cdots + u_i D^i+ \cdots
$$
and the corresponding infinite length codeword
$$
\vec v (D) = \vec v_0 + \vec v_1 D + \cdots + \vec v_i D^i+ \cdots, \mbox{where} \quad \vec v_i = (v_i^1,v_i^2)
$$
satisfy the matrix equation
\begin{equation}
\label{vuG}
    \vec v(D) = u(D)G(D).
\end{equation}
The transformation $u(D)$ to $\vec v(D)$ is invertible if
at least one of two polynomials is delay-free ($g_{10}=1$ or $g_{20}=1$).
The convolutional encoder (\ref{vuG}) is non-catastrophic if
$g_1(D)$ and $g_2(D)$ are relatively prime (\cite{RolfKam}). Since for any convolutional
code both catastrophic and non-catastrophic encoders do exist, we can consider
only non-catastrophic encoders without loss of optimality.

For any polynomial $a(D) = a_0 + a_1 D + \cdots + a_i D^i+ \cdots$
let $\lceil a(D) \rceil ^n = a_0 + a_1 D + \cdots + a_n D^n$
denote the finite degree polynomial obtained from $a(D)$ by truncating it to degree $n$.

Furthermore we consider $q$-ary linear $(n,k=n/2)$
codes which codewords are coefficients of truncated vector
polynomials $\lceil \vec v(D) \rceil ^{n/2}$ obtained from
degree $k=n/2$  information polynomial $u(d)$ by encoding
according to (\ref{vuG}).

\begin{figure}[ht]
\begin{minipage}{1\linewidth}
\begin{center}
\includegraphics[width=0.5\textwidth]{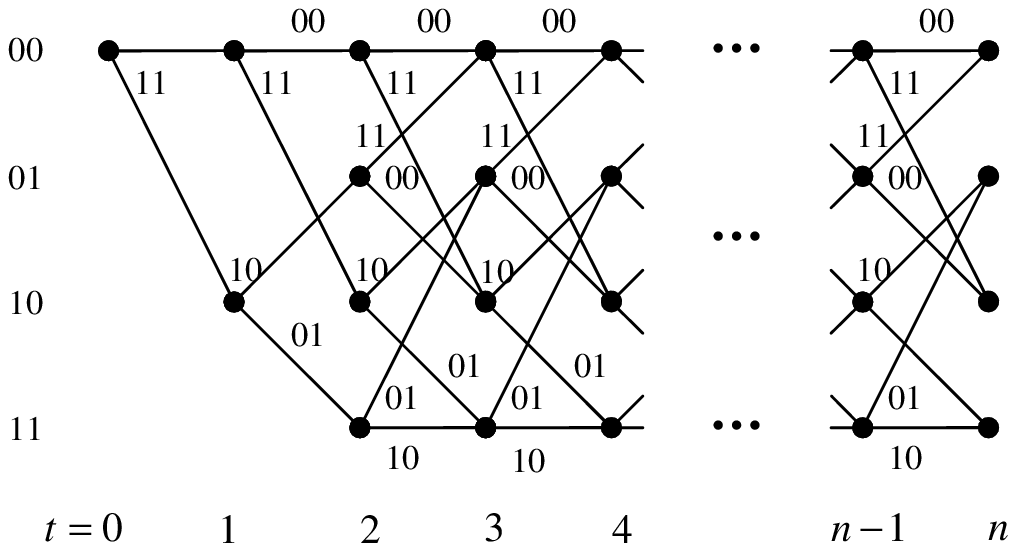}
\caption{Trellis representation of $(n,n/2)$ binary linear code obtained
from the $(1+D^2,1+D+D^2)$ convolutional encoder }
\label{conv_trellis}
\end{center}
\end{minipage}
\end{figure}

The trellis representation of $(n,n/2)$ binary linear code obtained
from $(1+D+D^2,1+D^2)$ convolutional encoder is shown in
Fig. \ref{conv_trellis}. For $q$-ary codes, $q>2$, similar trellises
can be constructed but with $q$ branches leaving each node and
merging  in each node of the trellis diagram.

To find the NSM for
a  $(g_1(D), g_2(D))$ convolutional encoder we first estimate the quantization
error for source sequences uniformly distributed in the hypercube $[0,q]^n$,
when the ``cyclic metric'' $\rho(\cdot,\cdot)$ (\ref{rho}) is
used as the distortion measure (see \cite{Kudryashov&Yurkov_PIT_2007}).
Then this error is recalculated into the NSM value using
(\ref{DistortionInHyperCube}).

To speed up the exhaustive search for the optimum
encoders over the set of all memory $m$ encoders with generators $(g_1(D),g_2(D))$
we used  the following rules to reduce the search space without loss of optimality,
\begin{itemize}
\item Only delay-free polynomials are considered $\displaystyle g^1_0 = g^2_0 = 1$.
\item Since the codes with the generators $(g_1(D),g_2(D))$ and $(g_2(D),g_1(D))$
 are equivalent, we consider only encoders with
  $\displaystyle \deg({g}_1(D)) \geq \deg({g}_2(D))$.
  If $\deg({g}_1(D)) = \deg({g}_2(D))$ then we require
  $|g_1(D)| \geq |g_2(D)|$, where we denote
  \[
  |a(D)|=\sum_{i=0}^{\deg(a(D))}a_i q^i.
  \]
In other words, we interpret the polynomial coefficients as positions of a number written in
$q$-ary form. Thereby we we assign an integer number (value) to each polynomial and
only pairs of generators sorted in descending order of the corresponding values
are used as candidates.
\item Only non-catastrophic encoders are considered, i.e.,
$\displaystyle \gcd(g_1(D), g_2(D))= 1$.
 \end{itemize}

\begin{table}
\renewcommand{\arraystretch}{1.3}
\caption{Optimum binary convolutional codes and their NSMs.
The generators are given in octal notation.}
\label{table_example2}
\begin{center}
\begin{tabular}{|c|c|c|c|}
\hline
states num. & generator & $G_n$ & En. gain, dB \\
\hline
2 & [3;1] & 0.0733 & 0.5571 \\
\hline
4 & [7;5] & 0.0665 & 0.9800 \\
\hline
8 & [17;13] & 0.0652 & 1.0657 \\
\hline
16 & [31;23] & 0.0643 & 1.1261 \\
\hline
32 & [61;57] & 0.0634 & 1.1873 \\
\hline
64 & [165;127] & 0.0628 & 1.2286\\
\hline
128 & [357;251] & 0.0623 & 1.2633\\
\hline
256 & [625;467] & 0.0620 & 1.2843\\
\hline
512 & [1207;1171] & 0.0618 & 1.2983\\
\hline
$\infty$ & --- & 0.0598 & 1.4389\\
\hline
\end{tabular}
\end{center}
\end{table}

\begin{table}
\renewcommand{\arraystretch}{1.3}
\caption{Optimum ternary convolutional codes and their NSMs.}
\label{table_example3}
\begin{center}
\begin{tabular}{|c|c|c|c|}
\hline
states num. & generator & $G_n$ & En. gain, dB \\
\hline
3 & [12;11] & 0.0720 & 0.6349 \\
\hline
9 & [121;111] & 0.0663 & 0.9931 \\
\hline
27 & [1211;1112] & 0.0641 & 1.1396 \\
\hline
81 & [11222;10121] & 0.0626 & 1.2424 \\
\hline
243 & [110221;101211] & 0.0617 & 1.3053 \\
\hline
729 & [1000112;112122] & 0.0614 & 1.3265\\
\hline
$\infty$ & --- & 0.0586 & 1.5231\\
\hline
\end{tabular}
\end{center}
\end{table}

\begin{table}
\renewcommand{\arraystretch}{1.3}
\caption{Optimum quinary convolutional codes and their NSMs.}
\label{table_example5}
\begin{center}
\begin{tabular}{|c|c|c|c|}
\hline
states num. & generator & $G_n$ & En. gain, dB \\
\hline
5 & [14;13] & 0.0716 & 0.6591 \\
\hline
25 & [131;102] & 0.0642 & 1.1328 \\
\hline
125 & [1323;1031] & 0.0622 & 1.2703 \\
\hline
625 & [10314;10133] & 0.0613 & 1.3336 \\
\hline
$\infty$ & --- & 0.0585 & 1.5229 \\
\hline
\end{tabular}
\end{center}
\end{table}

Our search results are presented in Tables \ref{table_example2},
\ref{table_example3}, and \ref{table_example5} for
$q=$2,3, and 5, respectively. It is easy to see that the NSM values of the optimum
codes decrease when $q$ grows, but the improvement provided by
ternary or quinary codes over binary codes is small and does not
worth the complication of the encoding.
Notice also, that the determined NSM values correspond to less shaping
gain than 1.36 dB reported in \cite{Marcellin&Fischer_1990} for
the 256-states quaternary codes. To achieve the same shaping
gain we have to have NSM=0.0609 whereas the best found 625-state
5-ary code provides NSM=0.0613 which is approximately 0.02 dB worse
than the value for the Marcellin-Fischer code.

Notice, however, that the Marcellin-Fischer quantizer  \cite{Marcellin&Fischer_1990}
is not a \emph{lattice} quantizer. Therefore, the NSM values reported here are the best among known multidimensional \emph{lattices}.

\section{Arithmetic coded lattice quantization}

Let us consider the lattice over a linear $(n,k)$ code $C$ (not necessarily a truncated
convolutional code). The entropy quantization
of the input sequence $\vec x$ can be performed in two steps:
\begin{itemize}
  \item Approximation, i.e., finding the vector of indices $\vec b$ and the codeword number
  $m\in \{0,...,q^k-1\}$ that minimizes the distortion value
  $d(\vec x, \vec \lambda)=n^{-1}\|\vec x - \vec \lambda \|^2=n^{-1}\|\vec x - (\vec c_m + q\vec b) \|^2$ (see (\ref{lattice point})).
  \item Entropy lossless coding of the pair (m,\vec b).
\end{itemize}

First, we consider the approximation step. From (\ref{lattice point}) the following reformulation
of the encoding problem follows,
\begin{eqnarray}
  \min_{\vec \lambda}  \left\{\|\vec x - \vec \lambda \|^2\right\}&=&
  \min_m \min_{\vec b}  \left\{\|\vec x - (\vec c_m + q\vec b) \|^2\right\}\nonumber\\
   &=& \min_m \min_{\vec b} \left\{ \sum_{t=1}^{n} (x_t - c_{mt} - qb_t) ^2\right\} \nonumber\\
   &=& \min_m \left\{\sum_{t=1}^{n}\min_{b_t} \left\{(x_t - c_{mt} - qb_t) ^2 \right\}\right\}\nonumber\\
   &=& \min_m \left\{\sum_{t=1}^{n}\mu(x_t, c_{mt})\right\} \nonumber\\
   &=& \min_m \left\{\mu(\vec x, \vec c_{m})\right\}.
   \label{approximation}
\end{eqnarray}
Here we introduced the additive metric
\begin{eqnarray}
    \mu(\vec x, \vec c) &=&\sum_{t=1}^{n}\mu(x_t, c_{t}), \nonumber\\
    \mu(x, c) &=& \min_{b}\left\{(x - c - q b) ^2\right\} = \left(x-c-q b_0 \right)^2\label{mu_metric}
 \end{eqnarray}
where the optimum index value $b_0$ can be determined by ``scalar quantization''
with step $q$ applied to $(x-c)$,
\begin{equation}
       b_0=\left\langle \frac{x-c}{q} \right\rangle. \label{b_optim}
\end{equation}
By $\langle \cdot \rangle$ we denote rounding to the nearest integer.

It follows from (\ref{approximation}), (\ref{mu_metric}), and (\ref{b_optim})
that lattice quantization can be split into two steps. First, for all
$t=1,...,n$ and $c=0,...,q-1$, the values $x_t-c$ should be scalar quantized and the $n\times q$ indices $b_{tc}$ and the corresponding metrics
$\mu_{tc}=(x_t - c - qb_{tc})^2$ are to be computed. In the second step, the codeword
$\vec c_{m}$
yielding the minimum of the metrics $\mu(\vec x, \vec c_{m})$ among all
codewords $\vec c_m\in C$, $m\in \{0,...,q^k-1\}$ should be found.
The codeword number $m$ and vector of indices $\vec b = (b_{tc_{m1}},...,b_{tc_{mn}})$
describe the trellis point closest to $\vec x$.
The formal description of the algorithm
is presented in Fig. \ref{algorithm_lattice_search}.

\begin{center}
\begin{figure}
\begin{algorithm}[H]
\dontprintsemicolon
    \KwIn { Source sequence $\vec{x}=(x_1,x_2,\ldots,x_n)$.  }
    \KwOut{ Number $m$ of codeword $\vec{c}_m\in C$,
    index sequence $\vec{b}=({b_1,b_2,\ldots,b_n})$. }
    \BlankLine
    \CommentSty{Metrics and indices computation:} \;
    \For {$t=1$ \emph{\KwTo} $n$}
    {
        \For {$c=0$ \emph{\KwTo} $q-1$}
        {
            $b_{t}(c)=\left\langle \frac{x_t-c}{q} \right\rangle$

            $\mu_t(c)=\left(x_t-c - q b_t(c) \right)^2$
        }
     }
\BlankLine
\CommentSty{Codeword search:}\;
Find the codeword $\vec{c}_m=(c_{m1}, c_{m2},\ldots, c_{mn})$ that minimizes metric
$\mu(\vec{x},\vec{c_m})= \sum_{t=1}^n \mu_t(c_{mt})$.
\BlankLine
\CommentSty{Result:}\;
The algorithm outputs the codeword number $m$ and the index vector $\vec{b}=(b_1(c_{m1}),b_2(c_{m2}),\ldots,b_n(c_{mn}))$.
\end{algorithm}
\caption{Approximation step of lattice quantization.}
\label{algorithm_lattice_search}
\end{figure}
\end{center}

\par The problem of searching for the optimum entry over the infinite cardinality
multidimensional lattice codebook is now divided into two steps:
$qn$ scalar quantization operations followed by searching
for the best codeword among the $q^k$ codewords of the $q$-ary linear $(n,k)$ code. The
latter step is similar to soft decision maximum-likelihood decoding of
error correcting linear block codes and there are many such decoding algorithms known;
in particular for truncated convolutional codes the Viterbi algorithm is a
reasonable choice.

Now let us consider an algorithm for a variable-length lossless coding
of the pair $(m,\vec b)$ obtained on the approximation step.

For a rate $R_{\rm C}=1/2$ truncated convolutional $(n,k=n/2)$-code, 
the codeword
is determined by length $k$ information sequence $\vec u =(u_1,...,u_k)$ and the codeword
can be represented as a sequence of $k$ tuples of length 2, i.e.,
$\vec c = (\vec c_1,..., \vec c_k)$, $\vec c_t=(c_t^1,c_t^2)$. We write the corresponding vector of
indices in tuples of length 2 too, i.e., $\vec b = (\vec b_1,..., \vec b_k)$, $\vec b_t=(b_t^1,b_t^2)$.

Denote by ${\vec \sigma_t}=(u_{t-1},...,u_{t-m})$  the encoder state at time $t$, where we
assume $u_t=0$ for $t\le 0$.

To describe  arithmetic coding (for details see \cite{WNC87}) it is enough to define a probability distribution on the pairs $(\vec u, \vec b)$. Thus we introduce the following two probability distributions,
\[
\{\vartheta(u_t|\vec \sigma_t)=\vartheta(u|\vec \sigma),
\quad \vec \sigma\in \mathbb{GF}_q^m \}
\]
\[
\{\varphi(b_{t}^i|c_{t}^i)=\varphi(b|c), b\in \mathbb{Z}, c \in\mathbb{GF}_q, i=1,2\}
\]
which we estimate for a given source pdf by training using a long simulated
data sequence. The required probability distribution
$p(\vec u, \vec b)$ is defined as
$$
p(\vec{u},\vec{b}) = \prod_{t=1}^{k} \vartheta(u_t|\sigma_t)   \varphi(b_{t}^1|c_{t}^1)
\varphi(b_{t}^2|c_{t}^2)
$$
where code symbols  $c_{t}^i, i=1,2$, are uniquely defined by $u_t$ and $\sigma_t$.

Notice that similar entropy coding for trellis quantization was described in \cite{JoshiCrumpFischer95},
\cite{Marcellin&Fischer_1990}, and \cite{Fischer&Yang_1998}. The difference is that
in these papers the approximation alphabet is finite, whereas in our algorithm the approximation alphabet
is not necessary finite since we allow any lattice point to be an approximation vector.

Simulation results of the described arithmetic-coded quantization will be presented in next section.

Below we present a modification which is efficient for low quantization rates for non-uniform probability
distributions.

As it was mentioned in the Introduction, the codes which are good in terms of granular gain (NSM value) are not necessarily optimal for non-uniform distributions. The reason is that spheres are not necessarily good Voronoi regions in this case. Moreover, Voronoi regions are not optimal quantization cells in this case. To understand this better, we consider properties of optimal entropy-constrained scalar quantization.

First of all, notice that for a wide class of probability distributions the simulation results for optimal non-uniform scalar quantization are presented in \cite{Farvardin&Modestino_1984}. It is easy to verify that, for example, for GGD
random variables with parameter $\alpha \le 0.5$, for all bit rates  the best trellis quantizers loose with respect to optimum scalar quantizers \cite{Fischer&Yang_1998}. The analysis of optimum and near-optimum scalar quantizers from
\cite{Farvardin&Modestino_1984}, \cite{KudrPorovOh} show that:
\begin{itemize}
  \item For GGD random variables the optimum ECSQ quantizer has the
  larger quantization cells near zero and smaller cells for large
  values.
  \item The quantization thresholds are not equal to half of the sum
  of the neighboring approximation values. Therefore, the minimum MSE
  (or minimum Euclidean distance) quantizers
  are not optimum ECSQ quantizers.
  \item The quantizers which have all quantization cells of equal
  size except the one symmetrically located around zero
  (so called ``extended zero zone'' (EZZ)  quantizer), has performances very close
  to that of the optimum ECSQ quantizer .
\end{itemize}

It follows from these observations that good trellis quantizers should be
obtained as a generalization of the EZZ quantizer to the multidimensional case.
(Another method of extending the zero zone for trellis quantization,
so called New Trellis Source Code (NTSC), have been presented  in
\cite{Fischer&Yang_1998}).

\begin{center}
\begin{figure}
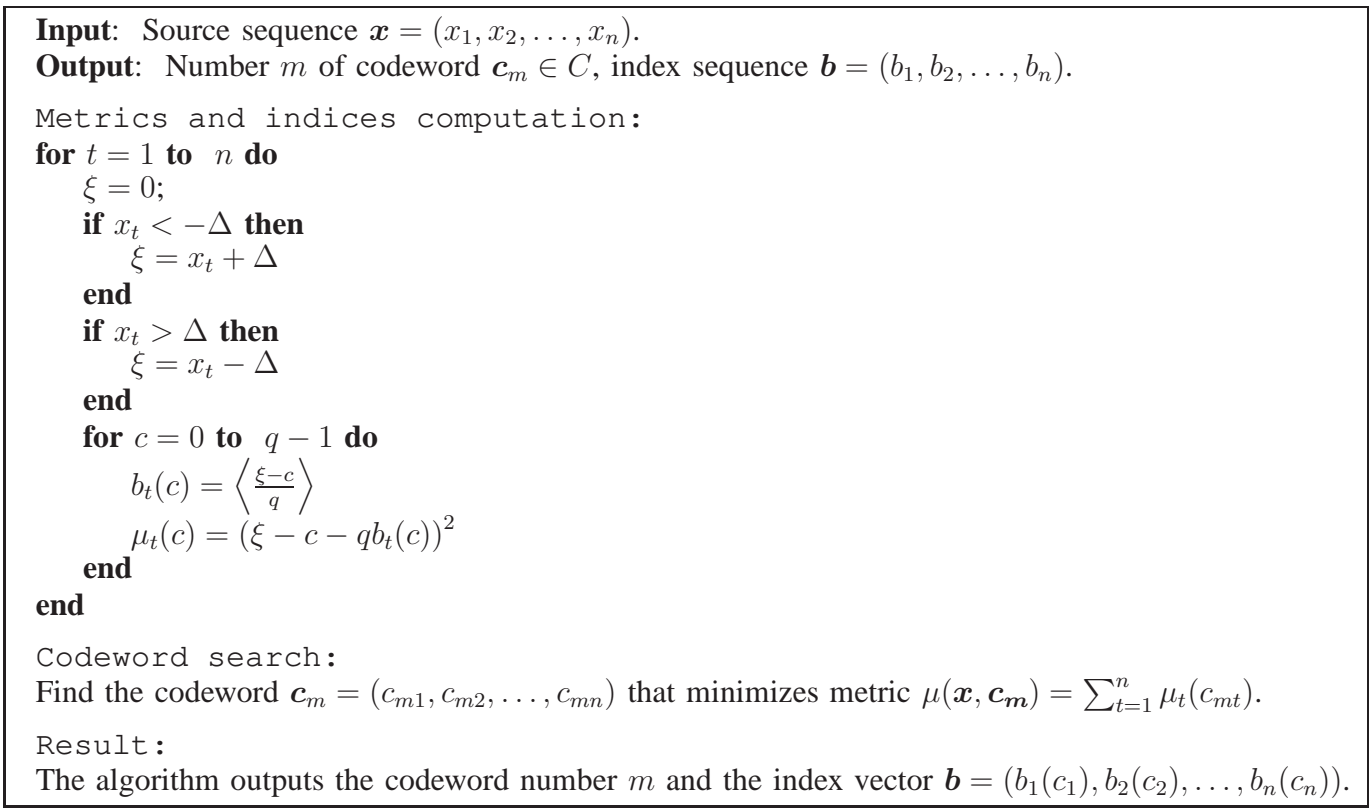

\begin{algorithm}[H]
\dontprintsemicolon
    \KwIn { Source sequence $\vec{x}=(x_1,x_2,\ldots,x_n)$.  }
    \KwOut{ Number $m$ of codeword $\vec{c}_m\in C$,
    index sequence $\vec{b}=({b_1,b_2,\ldots,b_n})$. }
    \BlankLine
    \CommentSty{Metrics and indices computation:} \;
    \For {$t=1$ \emph{\KwTo} $n$}
    { $\xi=0$;  \;
        \If {$x_t< -\Delta$} {$\xi = x_t+\Delta$}\;
        \If {$x_t>  \Delta$} {$\xi = x_t-\Delta$}\;
        \For {$c=0$ \emph{\KwTo} $q-1$}
        {
            $b_{t}(c)=\left\langle \frac{\xi-c}{q} \right\rangle$

            $\mu_t(c)=\left(\xi-c-q b_t(c) \right)^2$
        }
     }
\BlankLine
\CommentSty{Codeword search:}\;
Find the codeword $\vec{c}_m=(c_{m1}, c_{m2},\ldots, c_{mn})$ that minimizes metric
$\mu(\vec{x},\vec{c_m})= \sum_{t=1}^n \mu_t(c_{mt})$.
\BlankLine
\CommentSty{Result:}\;
The algorithm outputs the codeword number $m$ and the index vector $\vec{b}=(b_1(c_1),b_2(c_2),\ldots,b_n(c_n))$.
\end{algorithm}
\caption{Approximation with an extended zero zone}
\label{ezz_alg}
\end{figure}
\end{center}

To describe the multidimensional EZZ quantizer we need to generalize
(\ref{mu_metric}) and (\ref{approximation}) to the case when
the input data from some interval $[-\Delta, \Delta], \Delta>0$
are artificially interpreted as zeroes. Therefor
we replace each input r.v. $x$ by another  r.v. defined as
\begin{equation}\label{new_r_v}
    \xi=\left\{
    \begin{array}{ll}
      x+\Delta  &   x < -\Delta \\
      0,        & -\Delta \le x \le \Delta\\
      x-\Delta  & x > \Delta .
    \end{array}
    \right.
\end{equation}
Then we apply (\ref{mu_metric}) and (\ref{approximation}) to the r.v. $\xi$.
The generalized version of the algorithm describing the approximation step of
lattice quantization is shown in Fig. \ref{ezz_alg}.
The parameter $\Delta$ is different for different pdf's and quantization rates.
For each particular pdf and rate we minimize average distortion value
by selecting the best value of $\Delta$ from the set $2^{-s}, s=0,1,2,3$.
Simulation results will be presented in next section.

Now we consider the reconstruction of the approximation vector $\vec y =(y_1,...,y_n)$ by the
decoder from the received pair $(m, \vec b=(b_1,...,b_n))$. First, we introduce an additional
integer parameter $L$ which defines the amount of approximation values which are estimated
using a long training sequence (in our simulations we used $L=10$).
For the training source sequence $\vec x =(x_1,...,x_N)$ we performed the lattice quantization
described above and found the optimum codeword $\vec c =(c_1,...,c_N)$ and the corresponding
sequence of indices $\vec b =(b_1,...,b_N)$. Then for all $b \in \{-L,...,-1,1,...L\}$
and $c\in {0,\ldots,q-1}$ we computed ``refined'' approximation values
$$
\beta_{cb}=
\left\{
\begin{array}{ll}
\displaystyle 0, &  b=0,\quad c=0 \\
\displaystyle \frac{\sum_{i\in J_{cb}}x_i}{|J_{cb}|}+\Delta, & %
 b \in \{1,...L\},\quad c\in \{0,\ldots,q-1\}  \mbox{  or}\\
 & b=0,  \quad c\in \{1,\ldots,q-1\}\\
\displaystyle \frac{\sum_{i\in J_{cb}}x_i}{|J_{cb}|}-\Delta, &  b \in \{-L,...,-1\}, \quad c\in \{0,\ldots,q-1\}
\end{array}\right.
$$
where $J_{cb}=\{t|c_t=c,b_t=b\}$. The set of values $\beta_{cb}$ is assumed to be known to decoder.

When the pair $(m, \vec b=(b_1,...,b_n))$ is received, the decoder first finds the codeword
$\vec c_m =(c_{m1}...,c_{mn})$ and then reconstructs the approximation vector $\vec y =(y_1,...,y_n)$ as
$$
y_t=\left\{
\begin{array}{ll}
\beta_{c_{mt} b_t} & b_t \{-L,...,-1,1,...L\} \\
c_{mt}+qb_t   & b_t \notin \{-L,...,-1,1,...L\}.
\end{array}
\right.
$$
%
%
%

\section{Simulation results}

This section contains the results of the arithmetic coded lattice quantization for different types of sources. We consider the parametric class of generalized Gaussian distributions. The probability density function is parametrized by the parameter $\alpha$ and expressed as
$$
p(x)=\left[\frac{\alpha\eta(\alpha,\sigma)}{2\Gamma(1/\alpha)}\right]\exp\{-[\eta(\alpha,\sigma)|x|]^\alpha\}
$$
where
$$
\eta(\alpha,\sigma)=\sigma^{-1}\left[\frac{\Gamma(3/\alpha)}{\Gamma(1/\alpha)}\right]^{1/2}, \quad \alpha > 0
$$
\[
\Gamma(z)=\int_0^\infty t^{z-1}e^{-t}dt
\]
and $\sigma$ denotes the standard deviation. The Laplacian and Gaussian distributions are members of the family of generalized Gaussian distributions with parameters $\alpha=1.0$ and $\alpha=2.0$, respectively.

Simulation results for the Gaussian distribution are presented in Table \ref{Gauss}.
The SNR achieved by the ``New Trellis Source Code'' (NTSC)  \cite{Fischer&Yang_1998} and by
the ``Entropy Constrained Scalar Quantization'' (ECSQ) \cite{Farvardin&Modestino_1984} are also shown in the table. We can see that the estimated SNR value for the encoder with 512 states is only about 0.2 dB below
Shannon limit. For encoders with 4 and 32 states we obtain significant improvements over the NTSC-based quantization.

\begin{table}
\renewcommand{\arraystretch}{1.2}
\footnotesize
\caption{Gaussian source. Data for NTSC \cite{Fischer&Yang_1998} in parenthesis. }
\begin{center}
\begin{tabular}{|c|c|c|c|c|c|c|c|c|c|c|c|}
\hline
Rate & \multicolumn{9}{|c|}{Trellis states}  & $H(D)$ & ECSQ \\
\cline{2-10}
        & 2     & 4       & 8     & 16    & 32      & 64    & 128   & 256   & 512 &      &          \\
\hline
 0.5    &2.29&2.50    & 2.55  & 2.61  & 2.64    & 2.69  & 2.69  & 2.76  & 2.76  & 3.01  &  2.10 \\
\hline
 1      &5.06&  5.53   & 5.61  & 5.66  & 5.71    & 5.74  & 5.74  & 5.85  & 5.85  & 6.02  & 4.64  \\
        &    & (5.33)  &       &       & (5.64)  &       &       &       &       &       &       \\
\hline
 2      &11.08& 11.50   & 11.63 & 11.69 & 11.72   & 11.77 & 11.77 & 11.82 & 11.84 & 12.04 & 10.55 \\
        &     & (11.15) &       &       & (11.37) &       &       &       &       &       &       \\
\hline
 3      &17.11& 17.55   & 17.64 & 17.70 & 17.74   & 17.80 & 17.80 & 17.85 & 17.87 & 18.06 & 16.56 \\
        &     & (16.71) &       &       & (16.96) &       &       &       &       &       &       \\
\hline
\end{tabular}
\end{center}
\label{Gauss}
\end{table}

\begin{table}
\small
\renewcommand{\arraystretch}{1.3}
\caption{Laplacian source. Data for NTSC \cite{Fischer&Yang_1998} in parenthesis. }
\begin{center}
\begin{tabular}{|c|c|c|c|c|c|c|c|c|c|c|c|c|}
\hline
Rate &$\Delta$ &\multicolumn{9}{|c|}{Trellis states} & $H(D)$ & ECSQ \\
\cline{3-11}
      &  & 2     & 4       & 8     & 16    & 32      & 64    & 128   & 256   & 512   &      &       \\
\hline
 0.5& 0  & 2.92  & 3.03 & 3.06 & 3.09  & 3.11  & 3.14 & 3.17  & 3.18  & 3.21  &  3.54    & 3.11  \\
    & 0.25 & 3.06&3.15&3.17& 3.18&3.20 &3.22& 3.23&   &   & &   \\
\hline
 1 &0 &5.69 & 6.05 & 6.14  & 6.18  & 6.23   & 6.29 & 6.30 & 6.33  & 6.33  & 6.62 & 5.76 \\
&0.25 & 5.87   & 6.14  & 6.20  & 6.27  & 6.33    & 6.40  & 6.42  &   &  & &  \\
     &   &       & (5.93)  &       &       & (6.07)  &       &       &       &       &      &      \\
\hline
 2&0& 11.68& 12.15 & 12.22 & 12.29 & 12.32   & 12.38 & 12.41 & 12.43 & 12.44 & 12.66& 11.31\\
    &0.125&11.74 & 12.15   & 12.24 &  &  &  &  &  &  & & \\
    &    &       & (11.49) &       &       & (11.72) &       &       &       &       &      &      \\
\hline
 3 &0     & 17.74 & 18.15   & 18.23 & 18.29 & 18.33   & 18.39 & 18.41 & 18.44 & 18.44 & 18.68& 17.20\\
   &     &       & (17.00) &       &       & (17.15) &       &       &       &       &      &      \\
\hline
\end{tabular}
\end{center}
\label{Laplace}
\end{table}

\begin{table}
\small
\renewcommand{\arraystretch}{1.3}
\caption{Generalized Gaussian source with $\alpha = 0.5$. Data for NTSC \cite{Fischer&Yang_1998} in parenthesis. }
\begin{center}
\begin{tabular}{|c|c|c|c|c|c|c|c|c|c|c|c|c|}
\hline
Rate &$\Delta $& \multicolumn{5}{|c|}{Trellis states} & $H(D)$ & ECSQ \\
\cline{3-7}
    &    & 2     & 4       & 8     & 16    & 32      &       &      \\
\hline
 0.5 &  0    & 4.74  & 4.81    & 4.83  & 4.83& 4.84    &  5.62 & 5.37 \\
     &0.5  & 5.19  & 5.22    & 5.23  & 5.23  & 5.24    &  &  \\
\hline
 1 & 0    & 8.00  & 8.13  & 8.18  & 8.22  & 8.22    &  9.21 & 8.61 \\
   & 0.5  & 8.53  & 8.53  & 8.62  & 8.64  & 8.64    &   &  \\
   &     &       & (8.11)  &       &       & (8.29)  &       &      \\
\hline
 2 &  0    & 14.31 & 14.71  & 14.80& 14.90& 14.94  & 15.60 & 14.58\\
   &  0.25 & 14.56 & 14.90  & 15.03& 15.10& 15.14  &  & \\
   &       &       & (14.13) &       &       & (14.47) &       &      \\
\hline
 3 &  0  & 20.54 & 20.96 & 21.04& 21.10& 21.16   & 21.70 & 20.49\\
   & 0.25& 20.62 & 21.06  & 21.14& 21.20& 21.26  &       & \\
   &     &       & (19.61) &       &       & (19.92) &       &      \\
\hline
\end{tabular}
\end{center}
\label{GGD}
\end{table}

Simulation results for the Laplacian distribution and for the GGD with $\alpha=0.5$ are presented in
Tables \ref{Laplace} and \ref{GGD}, respectively. We show the quantization efficiency both without EZZ ($\Delta=0$) and with EZZ ($\Delta>0$). Also data from \cite{Fischer&Yang_1998} are given in
parenthesis. It follows from the simulation results that the EZZ is efficient for low quantization rates (below 2 bit/sample) and provides an SNR gain of about 0.1 dB for the Laplacian source and about 0.4 dB for the GGD with $\alpha=0.5$. The gap between the achieved SNR and the theoretical limit $H(D)$ for these two
probability distributions is nearly the same: 0.2 -- 0.3 dB.
Notice also, that for the GGD with $\alpha=0.5$, low-complexity (less than 32
encoder states) low rate quantizers do not perform better than the ECSQ. Therefore,
constructing efficient quantization algorithms for low rate coding
for the GGD with small $\alpha$ remains to be an open problem.


\section*{Appendix}
{\bf Comments to formulation of Theorem \ref{Main Teorem}.}

In \cite{Kudryashov&Yurkov_PIT_2007} expressions (\ref{d_0 calculation}) and
(\ref{R_0 calculation}) were given in the following form
\begin{equation}
\label{d_0 calculation_old}
d=\frac{1}{q}\left.\int_0^{q} \frac{\displaystyle \frac{\partial}
{\partial s}g(s,x)}{g(s,x)}dx\right|_{s=-\frac{1}{2d}}
\end{equation}
\begin{equation}
\label{R_0 calculation_old}
R_0(q)=\frac{1}{\ln q}
\Bigl(
-\frac{1}{2}-\frac{1}{q}\int_0^{q}
\ln g\left( 1/(2d_0),x \right) dx
\Bigr).
\end{equation}
The simplification is due to the observation that
$g(s,x)$ is a periodical function of $x$ with period 1 and inside the interval $[0,1]$ the function
$g(s,x)$ is symmetric with respect to the middle point $x=1/2$, i.e., $g(s,x)=g(s,1-x)$.

{\bf Proof of Corollary \ref{Cor_1}.}

Here we analyze the asymptotic behavior of the estimate of $G_n(q)$ given by Theorem \ref{Main Teorem}
when $q\rightarrow \infty$.

From 
definitions (\ref{rho}) and
(\ref{gsx}) for $x\in [0,0.5]$ we obtain
\begin{eqnarray}
\rho(x, k) &=& \left\{
\begin{array}{ll}
(x-k)^2, & k = 0,...,\frac{q-1}{2} \\
(k-x-q)^2, & k = \frac{q+1}{2},...,q-1
\end{array}\right.  \\
g(s,x)  & = & \frac{1}{q}\sum_{k=0}^{\frac{q-1}{2}}e^{s(x-k)^2}+\frac{1}{q}\sum_{k=\frac{q+1}{2}}^{q-1}e^{s(k-x-q)^2}\\
        & = & \frac{1}{q}\sum_{k=-\frac{q-1}{2}}^{\frac{q-1}{2}}e^{s(x-k)^2}, x\in [0,0.5].
\end{eqnarray}

To prove the Corollary, we choose
\begin{equation}
\label{sopt}
s=-\frac{\pi e}{q}
\end{equation}
The \emph{first} step is to verify that for large $q$ this
choice makes (\ref{d_0 calculation}) valid, i.e.,
\begin{equation}
\label{dopt}
d_0=-\frac{1}{2s}=\frac{q}{2\pi e}.
\end{equation}
The \emph{second} step is to substitute this expression into (\ref{R_0 calculation})
and show that for large $q$ we have $R_0(q)\to 1/2$. After these two steps we immediately
obtain the main result,
\[
\lim_{q \to \infty } G_\infty(q) = \lim_{q \to \infty } d_0 q^{2(R_0(q)-1)}=
\frac{1}{2\pi e}.
\]

We start the first step of derivations by demonstrating that asymptotically for large
$q$ the generating function $g(s,x)$ does not depend on $x$.
By straightforward computations it is easy to verify that
\begin{eqnarray}
\label{max_g_s_x}
\max_{x} \left\{ g(s,x) \right\}&=& g(s,0) = \frac{1}{q}\sum_{k=-\frac{q-1}{2}}^{\frac{q-1}{2}}e^{sk^2}  \\
\label{min_g_s_x}
\min_{x} \left\{ g(s,x) \right\}&=& g(s,1/2)=\frac{1}{q}\sum_{k=-\frac{q-1}{2}}^{\frac{q-1}{2}}e^{s(2k-1)^2/4}.
\end{eqnarray}
In order to estimate these sums we will use \cite{Dwight_eng_1961}, formula 552.6, that is,
\begin{equation}
\label{exp_dwight}
e^{ - x^2} + e^{ - 2^2x^2} + e^{ - 3^2x^2} + ... =
 - \frac{1}{2} + \frac{\sqrt \pi }{x}\left[ {\frac{1}{2} + e^{ - \pi ^2 /
x^2} + e^{ - 2^2\pi ^2 / x^2} + e^{ - 3^2\pi ^2 / x^2} + ...} \right].
\end{equation}

For large $q$ we obtain from (\ref{max_g_s_x}) that
\begin{eqnarray*}
g(s,0) & \le & \frac{1}{q} + \frac{2}{q}\sum_{k=1}^{\infty}e^{sk^2} \\
       & = & \frac{1}{q} + \frac{2}{q} \left(- \frac{1}{2} + \sqrt{-\frac{\pi }{s} }\left[ {\frac{1}{2} + e^{ \pi ^2 /
s} + e^{ 2^2\pi ^2 / s} + e^{ 3^2\pi ^2 / s} + ...} \right]\right).
\end{eqnarray*}

Substituting (\ref{sopt}) and upper-bounding the sum by the geometric progression for $q\ge 2$,
we obtain the upper bound
\begin{equation}
\label{upperbound}
g(s,x)\le \max_{x} g(s,x) \le \frac{1}{\sqrt{qe}} + e^{-q}.
\end{equation}

To obtain a lower bound on the sum in (\ref{min_g_s_x}) we rewrite it in the form
\[
g(s,1/2) =   \frac{2}{q}\sum_{k=1}^{\frac{q-1}{2}}e^{s(2k-1)^2/4}+\frac{1}{q}e^{sq^2/4}.
\]
Notice that the integer $(2k-1)$ in the exponent of summands runs over odd values. If we
add terms corresponding to intermediate even values the sum increases less than twice.
Therefore,
\[
g(s,1/2)  \ge   \frac{1}{q}\sum_{k=1}^{q}e^{sk^2/4}=
\frac{1}{q}\sum_{k=1}^{\infty}e^{sk^2/4}-\frac{1}{q}\sum_{k=q+1}^{\infty}e^{sk^2/4}.
\]
Next we estimate second sum by the geometric progression and estimate first sum using (\ref{exp_dwight}).
After straightforward computations we obtain
\begin{equation}
\label{lowerbound}
g(s,x)\ge \min_{x} g(s,x) \ge \frac{1}{\sqrt{qe}} - \frac{1}{q}.
\end{equation}

We can substitute (\ref{upperbound}) and (\ref{lowerbound}) into
(\ref{d_0 calculation}) to obtain lower and upper bounds on $d_0(q)$.
In both cases we have to estimate the following integral

\begin{eqnarray*}
&&2\int_0^{1/2} \frac{\partial}
{\partial s}g(s,x)dx =\\
&&\int_{-1/2}^{1/2}\frac{1}{q}\sum_{k = -\frac{q-1}{2}}^{\frac{q-1}{2}}(x-k)^2e^{s(x-k)^2}dx =\\
&&\frac{1}{q}\sum_{k = -\frac{q-1}{2}}^{\frac{q-1}{2}}\int_{-k-1/2}^{-k+1/2}x^2e^{sx^2}dx = \\
&&\frac{1}{q}\int_{-q/2}^{q/2}x^2e^{sx^2}dx = \\
&&\frac{\sqrt{q}}{2\pi e^{3/2}}{\rm erf}\left(\frac{\sqrt{\pi e q}}{2}\right) -
\frac{q^2}{4\pi e}e^{-\frac{\pi e q}{4}}.
\end{eqnarray*}
From inequalities (\cite{Dwight_eng_1961}, formula 592)
\[
1-\frac {e^{-x^2}}{x\sqrt{\pi}}\le {\rm erf}(x)\le 1
\]
follow the bounds
\[
\frac{\sqrt{q}}{2\pi e^{3/2}} -
\frac{q^2}{2\pi e}e^{-\frac{\pi e q}{4}}
\le 2\int_0^{1/2} \frac{\partial}{\partial s}g(s,x)dx \le \frac{\sqrt{q}}{2\pi e^{3/2}}.
\]
Substituting them together with
(\ref{upperbound}) and (\ref{lowerbound}) into (\ref{d_0 calculation})
we obtain
\[
\frac{q}{2\pi e}\left(1-O\left(q^2e^{-q} \right)\right)
\leq d_0 \leq
\frac{q}{2\pi e}\left(1+O\left(\frac{1}{\sqrt{q}} \right)\right).
\]
Therefore (\ref{dopt}) is proven and the first step completed.

To fulfil the second step and the proof of Corollary, we have to substitute the bounds (\ref{upperbound}) and (\ref{lowerbound}) into expression (\ref{R_0 calculation}). After simple derivations we immediately obtain  (\ref{Rinf}) \QED


\bibliographystyle{ieeetr}

\end{document}